\documentclass[reqno]{amsart}
\usepackage{amssymb,amsthm,amsfonts,amstext}
\usepackage{amsmath}

\makeatletter \@addtoreset{equation}{section} \makeatother

\renewcommand\thefigure{\thesection.\@arabic\c@figure}
\renewcommand\thetable{\thesection.\@arabic\c@table}
\newtheorem{theorem}{Theorem}[section]

\newtheorem{proposition}[theorem]{Proposition}

\newcommand{\mc}[1]{{\mathcal #1}}

\newcommand{\mb}[1]{{\mathbf #1}}
\newcommand{\bb}[1]{{\mathbb #1}}

\newcommand{\<}{\langle}
\renewcommand{\>}{\rangle}

\title[Tagged particle in single-file diffusion]{Scaling limits of a tagged particle in the exclusion process with variable diffusion coefficient}

\begin{document}

\author{Patr\'{\i}cia Gon\c{c}alves}
\address{Centro de Matem\'atica \\
Campus de Gualtar\\ 4710-057 Braga, Portugal}
\email{patg@impa.br}
\author{ Milton Jara }
\address{FYMA\\ Universit\'e Catholique de Louvain\\ 2, Chemin du Cyclotron, B-1348\\ Louvain-la-Neuve, Belgium}
\email{milton.jara@uclouvain.be}

\date{\today}

\begin{abstract}
We prove a law of large numbers and a central limit theorem for a tagged particle in a symmetric simple exclusion process in $\bb Z$ with variable diffusion coefficient. The scaling limits are obtained from a similar result for the current through $-1/2$ for a zero-range process with bond disorder. For the CLT, we prove convergence to a fractional Brownian motion of Hurst exponent $1/4$.
\end{abstract}


\renewcommand{\subjclassname}{\textup{2000} Mathematics Subject Classification}

\maketitle

\section{Introduction}

A classical problem in statistical mechanics consists in obtaining the scaling limit of a tagged (sometimes called tracer) particle in a system of interacting particles. In general, we expect the limiting process to be a Brownian motion for the system in equilibrium \cite{KV}, or a time-inhomogeneous diffusion in general \cite{JLS}. However, in the special case of a one-dimensional, nearest-neighbors system (the so-called {\em single-file diffusion}), a sub-diffusive behavior is expected \cite{Har}. In the case of a symmetric, simple exclusion process, the scaling limit of a tagged particle is given by $\sqrt{(1-\varphi)/\varphi)} \mc Z_t$, where $\mc Z_t$ is a fractional Brownian motion of Hurst exponent $1/4$ and $\varphi$ is the density of particles \cite{Arr}, \cite{DF}. 

In recent years, the evolution of a random walk in random environment have received a lot of attention. Therefore, to investigate the behavior of particle systems in random environments seems to be a natural step ahead. A simple example of a single-file diffusion with disorder is what we call the exclusion process with variable diffusion coefficient. In this model, to each particle we attach a different diffusion coefficient,  and we let the system evolve in such a way that each particle performs a continuous-time, simple random walk with its own diffusion coefficient, conditioned to have at most one particle per site (the {\em exclusion} rule). For a physical interpretation of this model and applications, see \cite{Asl}, \cite{ALS}.

In these notes we generalize the results of \cite{Arr}, \cite{DF} to the simple exclusion process with variable diffusion coefficient. We will see that the scaling limits depend on the diffusion coefficient only through an {\em effective} diffusion coefficient, obtained from the related homogenization problem. 
Due to the variable diffusion coefficient, the approach of \cite{Arr}, \cite{DF} can not be repeated here, since in the associated stirring process, particles not only exchange their positions, but also their diffusion coefficients. We adopt here the approach of \cite{LOV2}, \cite{LV}. There, scaling limits for the tagged particle are related to scaling limits for the current through the $\<-1,0\>$ bond in a {\em zero-range} dynamics. We call this relation the {\em coupling method}. The main problem is that the associated zero-range dynamics in the case of variable diffusion coefficient is a particle system with {\em bond disorder}. 
As proposed by Rost and Vares \cite{RV}, scaling limit for the current can be obtained from the hydrodynamic limit and density fluctuations of the density of particles.
To obtain those scaling limits in the presence of random media, various methods have been developed \cite{Fri2}, \cite{Qua2}, \cite{FM}. However, the  method developed recently in \cite{GJ} applies directly to our situation, avoiding the hard non-gradient tools needed in the other references. This is specially convenient in this setting, since the so-called spectral gap for the zero-range process in not uniform in density for our model. 
In fact, we think of this work as a nice direct application of the results in \cite{GJ}.

This article  is organized as follows. In section \ref{s1} we define the model and state the main results. In section \ref{s2} we construct the associated zero-range dynamics and we prove the main results. In section \ref{s3} we discuss the relevance of the assumptions made on initial conditions and on the variable diffusion coefficients.

\section{Definitions and results}
\label{s1}

Consider a system of continuous-time, interacting random walks $\{x_i(t); i \in \bb Z\}$ on the integer lattice $\bb Z$, and a sequence of positive numbers $\{\lambda_i;i \in \bb Z\}$. The dynamics of this system is the following. The particle $x_i(t)$ waits an exponential time of rate $\lambda_i$, at the end of which it chooses one of its two neighbors with equal probability. If the chosen site is empty, the particle jumps to that site; if the site is occupied for another particle, the jump is not realized and a new exponential time starts afresh. We call this interaction the {\em exclusion} rule. This dynamics corresponds to a Markov process defined on the state space $\Omega^{ex}_0=\{\mathbf x \in \bb Z^{\bb Z}; x_i \neq x_j \text{ for } i\neq j\}$ and generated by the operator
\begin{align*}
L^{ex}_0 f(\mb x) = \sum_{i \in \bb Z} \lambda_i\big\{\mb 1(\mb x +e_i\in \Omega_0^{ex})\big[f(\mb x+e_i)-f(\mb x)\big] \\
	+\mb 1(\mb x -e_i\in \Omega_0^{ex})\big[f(\mb x-e_i)-f(\mb x)\big]\big\}. 
\end{align*}

The number $\lambda_i$ is interpreted as the {\em diffusion coefficient} of particle $x_i$. 
We call the system $\{x_i(t)\}_i$ described in this way, the simple exclusion process with variable diffusion coefficient, and we call $\{\lambda_i\}_i$ the diffusion coefficient of $\{x_i(t)\}$. Notice that when $\lambda_i = \lambda$ does not depend on $i$, the process $\{x_i(t)\}$ corresponds to a labeling of  the usual simple exclusion process. We refer to this situation as the homogeneous case. It will be of interest to define the particle configuration $\eta_t \in \Omega^{ex}=\{0,1\}^{\bb Z}$ by $\eta_t(x) = 1$ if $x_i(t)=x$ for some $i \in \bb Z$. Notice that $\eta_t$ is {\em not} a Markov process, since in this case particles are indistinguishable, and therefore we can not keep track of the different diffusion coefficients. It will be useful as well to define the {\em environment process} $\eta^{env}_t(x) = \eta_t(x+x_0(t))$. Differently from the process $\eta_t$, $\eta^{env}_t$ is indeed a Markov process. In fact, due to the exclusion rule and the nearest-neighbor jumps, the relative ordering of particles is preserved by the dynamics. From now on, we assume, relabeling the particles if needed, that $x_i(t)< x_{i+1}(t)$ for any $i \in \bb Z$. Therefore, in the process $\eta^{env}(t)$, the $i$-th particle at the right of the origin has diffusion coefficient $\lambda_i$, and analogously for particles to the left of the origin.

Our aim is to study the asymptotic behavior of a tagged particle for this system. Let us assume that $x_0(0)=0$, that is, that the $0$-th particle starts at the origin. In the homogeneous case, it is well known that the behavior of a tagged particle is sub-diffusive, with a scaling exponent equal to $1/4$ (instead of the diffusive exponent $1/2$). We will see that this is also the case for the exclusion process with variable diffusion coefficient. For simplicity, we will assume that the sequence $\{\lambda_i\}_i$ is an i.i.d sequence of random variables satisfying the ellipticity condition $\epsilon_0 \leq \lambda_i \leq \epsilon_0^{-1}$ for any $i \in \bb Z$. More general conditions on $\{\lambda_i\}_i$ will be discussed below. 

Now we explain how do we construct the initial configuration for the process $\{x_i(t)\}$.
Let $u_0: \bb R \to [0,1]$ be a continuous function. For each $n >0$, we define the initial distribution $\mu_n$ as the product measure in $\Omega^{ex}$ with marginals
\[
\mu_n\{\eta(x)=1\}=
\begin{cases}
u_0(x/n), &x \neq 0\\
1, &x=0.\\
\end{cases}
\]

We define $x_0(0)=0$. For $i>0$, we define $x_i(0)$ recursively by $x_i(0)= \inf\{x>x_{i-1}(0); \eta(x)=1\}$. For $i<0$, we adopt a similar procedure. We denote this random initial configuration by $\mb x^n_{u_0}$. Now we are ready to state our first result, which is a law of large numbers for $x_0(t)$. For technical reasons, we will restrict ourselves to the case on which $\lim_{x \to \pm \infty} u_0(x) = u_0(\pm \infty)$ both exist and are different from $0,1$ and on which $\sup_x u_0(x) <1$.

\begin{theorem}
\label{t1}
Let $\{x_i(t)\}$ be the exclusion process with variable diffusion coefficient $\{\lambda_i\}_i$ and starting from $\mb x_{u_0}^n$. For almost every realization of the environment $\{\lambda_i\}_i$,  
\[
\lim_{n \to \infty} \frac{x_0(tn^2)}{n} = r(t)
\]
in probability, where $r(t)$ is the solution of 
\[
\int_0^{r(t)} u(t,x) dx = -\int_0^t \partial_x u(s,0) ds,
\]
$u(t,x)$ is the solution of the hydrodynamic equation $\partial_t u= \lambda \partial^2_x u$ with initial condition $u_0$ and $\lambda = (E[\lambda_i^{-1}])^{-1}$.	
\end{theorem}

Notice in the previous theorem, the subdiffusive scaling, the dependence on the {\em effective} diffusion coefficient $\lambda$ and the {\em quenched} statement about the random diffusion coefficients. 

Once we have obtained a law of large numbers, a natural question is whether a central limit theorem holds. 

\begin{theorem}
\label{t2}
Consider an initial flat profile $u_0(x) \equiv \varphi$.
Let $\{x_i(t)\}$ be the exclusion process with variable diffusion coefficient $\{\lambda_i\}_i$ and starting from $\mb x_{\varphi}^n$. Then, 
\[
\lim_{n \to \infty} \frac{1}{n^{1/2}} x_0(tn^2) = \sqrt{\frac{(1-\varphi)\sqrt{\lambda}}{\varphi}} \mc Z_t,
\]
where $\mc Z_t$ is a symmetric fractional Brownian motion of Hurst index $1/4$ and the convergence is in the sense of finite distributions convergence.  
\end{theorem}

Notice in the previous theorem the explicit dependence of the limiting process in $\lambda$ and $\varphi$.

\section{The coupling method}
\label{s2}

As we mentioned before, the environment process $\eta_t^{env}$ is a Markov process, but with a complicated non-local dynamics due to the ordering of particles. In order to obtain a local dynamics, we define a new process $\xi_t$ with state space $\Omega^{zr} = \bb N_0^{\bb Z}$ by taking $\xi_t(i) = x_{i+1}(t) - x_i(t)$. It is easy to verify that $\xi_t$ is a Markov process, generated by the operator
\[
L^{zr} f(\xi) = \sum_{i \in \bb Z} \lambda_i \Big\{ g\big(\xi(i)\big)\big[f(\xi^{i,i+1})-f(\xi)\big] + g\big(\xi(i+1)\big) \big[f(\xi^{i+1,i})-f(\xi)\big] \Big\},
\]
where $g:\bb N_0 \to \bb R$ is defined by $g(0)=0$, $g(i) =1$ if $i \geq 1$, and $\xi^{i,j}$ is defined by
\[
\xi^{i,j}(l) =
\begin{cases}
\xi(i)-1,& l=i\\
\xi(j)+1, &l=j\\
\xi(l), &l \neq i,j.\\
\end{cases}
\]

The process $\xi_t$ is known in the literature as a {\em zero-range process with bond disorder}. Although this functional transformation of the simple exclusion into a zero-range process was already exploited by Kipnis \cite{Kip} to obtain the scaling limit of a tagged particle in the asymmetric simple exclusion process, the mathematical tools in order to treat the zero-range process with bond disorder have been developed only recently \cite{GJ}, \cite{JL2}. The dynamics of this model can be described as follows. At each bond $\<i,i+1\>$ we attach a {\em conductance} $\lambda_i$, corresponding to a exponential clock of rate $\lambda_i$. Each time this clock rings, we choose one of the directions $i \to i+1$ or $i+1 \to i$ with equal probability. Then we displace a particle from the origin site to the destination site. If there is not particle on the origin site at that time, nothing happens. We do this independently for any bond $\<i,i+1\>$. 

Let us denote by $j_{x,x+1}(t)$ the total current of particles through the bond $\<x,x+1\>$ for the process $\xi_t$.
The key observation is that the position of the tagged particle $x_0(t)$ in the original model corresponds to the current through the bond $\<-1,0\>$ in the model $\xi_t$. Therefore, Theorem \ref{t1} is consequence of the following result:

\begin{theorem}
\label{t3}
Under the assumptions of Theorem \ref{t1}, 
\[
\lim_{n \to \infty} \frac{1}{n} j_{-1,0}(tn^2) = - \int_0^t \lambda \Phi\big(v(s,0)\big)\partial_x v(s,0)ds,
\]
where $v(t,x)$ is the solution of the Cauchy problem
\begin{equation}
\label{ec1}
\begin{cases}
\partial_t v(t,x) &=\partial_x \Big(\lambda\Phi\big(v(t,x)\big) \partial_x v(t,x)\Big)\\
v(0,x) &= v_0(x),
\end{cases}
\end{equation}
the function $\Phi(v)$ is given by $\Phi(v) = 1/(1+v)^2$ and the initial condition $v_0(x)$ is defined below.
\end{theorem}

About the initial condition $v_0$, we notice that our choice of $\mb x^n_{u_0}$ induces a choice $\xi^n_0$ in the initial distribution of $\xi_t$ as well. Therefore, we define $v_0$ as the functional transformation of $u_0$ in the continuous limit. This is given by the relation
\[
\int_0^{z+\int_0^z v(x) dx} u(y)dy =z.
\]

Defining $V(z) = z+ \int_0^z v_0(x)dx$, we see that $V$ satisfies the ODE
\[
\begin{cases}
\frac{d}{dz} V(z) &= u_0(V(z))^{-1}\\
V(0) &=0,
\end{cases}
\]
from where we obtain $v_0$ by taking $v_0(x) = V'(x)-1=\big(1-u_0(V(x))\big)/u_0(V(x))$.

In the same way, Theorem \ref{t2} is consequence of the following result:

\begin{theorem}
\label{t4}
Under the assumptions of Theorem \ref{t2}, 
\[
\lim_{n \to \infty} \frac{1}{n^{1/2}} j_{-1,0}(tn^2) = \sqrt{\frac{(1-\varphi)\sqrt{\lambda}}{\lambda \varphi}} \mc Z_t.
\]
\end{theorem}

\subsection{The law of large numbers}

Equation \ref{ec1} corresponds to the {\em hydrodynamic equation} of the process $\xi_t$. For the process $\xi_t$ restricted to a finite volume, the so-called hydrodynamic limit has been obtained in \cite{GJ} (see also \cite{JL2}). Under the technical condition $u(\pm \infty)\neq 0,1$, the result in \cite{GJ} can be extended to infinite volume as in \cite{BKL}. Therefore, we have the following 

\begin{proposition}
\label{p1}
For any continuous function $G: \bb R \to \bb R$ of bounded support, we have
\[
\lim_{n \to \infty} \frac{1}{n} \sum_{x \in \bb Z} \xi_{tn^2}(x) G(x/n) = \int_{\bb R} G(x) v(t,x) dx
\]
in probability with respect to the law of $\xi_t$ starting from $\xi^n_0$, where $v(t,x)$ is the solution of the hydrodynamic equation \ref{ec1}.
\end{proposition}

Starting from this proposition, the law of large numbers for $j_{-1,0}(t)$ is easy to understand. In fact, the macroscopic density flux across 0 in the hydrodynamic equation is given by $J_0(t)= -\int_0^t \lambda \Phi(v(s,0)) \partial_x v(s,0) ds$. Therefore, Theorem \ref{t3} states that the current $j_{-1,0}(t)$, when properly rescaled, converges to $J_0(t)$. Notice that Proposition \ref{p1} does not give the desired result immediately, since an exchange of limits between the derivative and the scaling has to be justified. In order to justify this exchange, we can proceed as in \cite{JL1}, so we will just give an outline of the arguments. When the total mass $\int v_0(x)dx$ of the initial condition $v_0(x)$ is finite, the integrated current satisfies
\begin{equation}
\label{ec2}
J_0(t) = \int_0^\infty \big\{v(t,x)-v(0,x) \big\} dx.
\end{equation}

The identity is just saying that the total current across 0 is equal to the mass at time $t$ to the right of the origin, minus the mass at the right of the origin at time $0$, an integrated version of the conservation of mass.
Let us define $G_0(x)= \mb 1\{x \geq 0\}$, the Heaviside function. We can rewrite $J_0(t) = \int_{\bb R} G_0(x) \big(v(t,x) -v(0,x)\big) ds$ and in an analogous way, 
\begin{equation}
\label{ec3}
\frac{1}{n} j_{-1,0}(tn^2) = \frac{1}{n} \sum_{x \in \bb Z} \big\{\xi_{tn^2}(x) -\xi_0(x)\big\} G_0(x/n).
\end{equation}

If the initial profile has non-null asymptotic densities $v_0(\pm\infty) = \lim_{x \to \pm \infty} v_0(x)$, then formula (\ref{ec2}) still makes sense, but we can not send $n$ to $\infty$ in (\ref{ec3}), since $G_0$ is not of compact support. Let us define $G_l(x) = (1-x/l)^+$ for $x \geq 0$ and $G_l(x)=0$ for $x<0$. Here $(\cdot)^+$ denotes positive part. Taking the Cauchy limit of the right side of the following expression, it is not difficult to see that 
\[
\frac{1}{n}j_{-1,0}(tn^2) = \lim_{l \to \infty} \frac{1}{n} \sum_{x \in \bb Z} \big\{\xi_{tn^2}(x) -\xi_0(x)\big\} G_l(x/l),
\]
the limit being in $\mc L^2(\bb P_n)$, where $\bb P_n$ is the law of the process $\xi_{tn^2}$ starting from $\xi_0^n$. Moreover, the limit is uniform in $l$. But now, since the support of $G_l$ is compact, we can use the result of Proposition \ref{p1} to pass to the limit in the previous expression, and to obtain the desired result.

\subsection{The central limit theorem} We recall the strategy in order to prove a law of large numbers for the current $j_{-1,0}(t)$ followed in the previous section. The idea is to relate the current to some linear functionals of the density of particles. This relation holds for the microscopic model, and also for the density flux in the macroscopic equation. In this way, the scaling limit of the current should follow from the scaling limit of the density of particles. However, the functional considered was not continuous, and therefore a cut-off argument was needed in order to take the limit. The final step was to justify an exchange of limits between the scaling and the cut-off. This strategy also gives a central limit theorem for the current in the zero-range process $\xi_t$. Differently from \cite{JL1}, in the case of the zero-range process the central limit theorem for the density of particles has only been obtained in equilibrium, restricting ourselves to that situation.

Now we explain what do we mean by a central limit theorem for the density of particles. For the initial configuration $\mb x^n_\varphi$, the associated initial distribution $\xi_0^n$ is a product, uniform measure $\bar \nu_\varphi$ in $\Omega^{zr}$ with marginals given by
\[
\bar \nu_\varphi(\xi(x)=k) = \varphi(1-\varphi)^k.
\]

A simple computation shows that the measure $\bar \nu_\varphi$ satisfies detailed balance with respect to the evolution of $\xi_t$, and therefore it is an invariant, reversible measure for this dynamics. It is not hard to check that it is also an ergodic measure for $\xi_t$. Let us notice that the number of particles in the zero-range process $\xi_t$ is locally conserved, and due to ergodicity, it is the only locally conserved quantity. Therefore, it is more natural to parametrize the family $\{\bar \nu_\varphi; \varphi \in [0,1]\}$ of ergodic, invariant measures by the density of particles at each site. It is easy to see that the density of particles $\rho(\varphi)$ defined by $\rho(\varphi) = \int \xi(x) \bar \nu_\varphi$ satisfies $\rho(\varphi) = (1-\varphi)/\varphi$. We define then $\varphi(\rho)= 1/(1+\rho)$ and $\nu_\rho = \bar \nu_{\varphi(\rho)}$. Now fix a density $\rho \in (0,\infty)$ and consider the process $\xi_t$, starting from the measure $\nu_\rho$. This is equivalent to start the exclusion process with variable diffusion coefficient from the configuration $\bf x^n_\varphi$ for $\varphi= 1/(1+\rho)$.

Let us define the $\mc S'(\bb R)$-valued process $\mc Y_t^n$ by
\[
\mc Y_t^n(G) = \frac{1}{\sqrt n} \sum_{x \in \bb Z} G(x/n) \big( \xi_{tn^2}(x)-\rho\big),
\]
where $G$ is in $\mc S(\bb R)$, the Schwartz space of test functions in $\bb R$. We call this process the {\em density fluctuation field}.
The following proposition has been proved in \cite{GJ}:

\begin{proposition}
\label{p2}
The fluctuation field $\mc Y_t^n$ converges in the sense of finite distributions to a generalized Ornstein-Uhlenbeck process of mean zero and characteristics $\lambda \Phi(\rho)\partial_x^2$ and $\sqrt{\lambda\rho/(1+\rho)} \partial_x$. 
\end{proposition}

As in the case of the law of large numbers, the current $j_{-1,0}(t)$ satisfies
\[
\frac{1}{\sqrt n} j_{-1,0}(tn^2) = \lim_{l \to \infty} \mc Y_t^n(G_l) - \mc Y_0^n(G_l),
\]
the limit being uniform in $\mc L^2(\bb P_n)$. Therefore, we conclude that
\[
\lim_{n \to \infty} 
\frac{1}{\sqrt n} j_{-1,0}(tn^2) = \lim_{l \to \infty} \mc Y_t(G_l) - \mc Y_0(G_l),
\]
from where Theorem \ref{t4} follows.

\section{Conclusions and remarks}
\label{s3}
\subsection*{1}  Comparing our results with the classical result of Arratia \cite{Arr} and the results in \cite{JL2}, we see that the variable diffusion coefficient enters into the scaling limit of the tagged particle only through the effective diffusion coefficient $\lambda$. Notice that this effective diffusion coefficient is the same obtained by homogenization, although our system is subdiffusive. This is due to the fact that, for particle systems with bond disorder, the disorder and the time evolution can be decoupled using an stochastic version of the compensated compactness lemma \cite{GJ}. We conclude that the subdiffusive behavior is a purely dynamical feature, since the factor $\sqrt{(1-\varphi)\sqrt{\lambda}/ \varphi}$ is the same appearing for a simple exclusion process with uniform diffusion coefficient $\lambda$. In particular, the variance of $x_0(t)$ is of order $\sqrt{2\lambda t/\pi} (1-\varphi)/ \varphi$ as $t \to \infty$.

\subsection*{2} About our assumptions on the initial conditions for the law of large numbers, the product structure of the initial measure $\mu_n$ is far from necessary. The more general assumptions are the following. First we need the initial measure $\mu_n$ to be {\em associated} to some macroscopic profile $u_0$, in the sense that the particle density should converge in distribution to $u_0(\cdot)$ (for a precise definition, see Section 2.3 of \cite{GJ}). The second assumption is a bound on the entropy density of the initial measure $\mu_n$ with respect to the measure $\mu^\varphi$ obtained as the product measure associated to the flat profile $u_0 \equiv \varphi$. We also need the distribution of $\xi_0^n$ to be stochastically dominated by the measure $\nu_\rho$ for some $\rho>0$. This imposes the restriction $\sup_x u_0(x) <1$.

\subsection*{3} In the CLT for the tagged particle, a key input is the fluctuation result for the particle density, Proposition \ref{p2}. A complete proof of the results in \ref{p2} is only available in equilibrium, making the choice of $\mathbf x^n_\varphi$ rather restrictive. To extend Proposition \ref{p2} to the non-equilibrium setting remains one of the main open problems in the theory of scaling limits of particle systems. Notice that such a result have been obtained in \cite{JL2} for the simple exclusion process in random media and 
in \cite{Jan} for the zero-range process. None of these approaches can be adapted to our situation, however.

\subsection*{4} The assumptions on the variable diffusion coefficients $\{\lambda_i\}_i$ are not, by far, optimal. We have taken $\{\lambda_i\}_i$ as an i.i.d. sequence just to fix ideas. In fact, in order to get the results of Theorem \ref{t1}, it is enough that $\sup_i \lambda_i >0$ and 
\[
\lim_{n \to \infty} \frac{1}{n} \sum_{i=-n}^0 \lambda_i^{-1} = \lim_{n \to \infty} \frac{1}{n} \sum_{i=0}^n \lambda_i^{-1} = \lambda^{-1}
\]
both exist and are different from 0. To obtain the results in Theorem \ref{t2}, we need in addition the ellipticity condition $\epsilon_0^{-1}>\lambda_i >\epsilon_0$ for every $i$. In particular, both results remain true if for instance we take a periodic sequence $\{\lambda_i\}_i$ of diffusion coefficients.

\section*{Acknowledgments}
M.J. was supported by the Belgian Interuniversity Attraction Poles Program P6/02, through the network NOSY (Nonlinear systems, stochastic processes and statistical mechanics). P.G. would like to thank the hospitality at Universit\'e Catholique de Louvain, where this work was finished.

\end{document}